\documentclass[aip,apl,reprint,groupedaddress,graphicx]{revtex4-1}
\usepackage{graphicx}
\usepackage{color}
\usepackage{dcolumn}
\draft 

\begin{document}
\title{Large magnetothermopower effect in Dirac materials (Sr/Ca)MnBi$_2$} 

\author{Kefeng Wang, Limin Wang and C. Petrovic}
\affiliation{Condensed Matter Physics and Materials Science Department, Brookhaven National Laboratory, Upton, New York 11973 USA}

\date{\today}

\begin{abstract}
We report temperature and magnetic field dependence of the thermal transport properties in single crystals of (Sr/Ca)MnBi$_2$ with linear energy dispersion. In SrMnBi$_2$ thermopower is positive, indicating hole-type carriers and the magnetic field enhances the thermopower significantly. The maximum change of thermopower is about $1600\%$ in 9 T field and at 10 K. A negative thermopower is observed in CaMnBi$_2$ with dominant electron-type carriers and, in contrast, the magnetic field suppresses the absolute value of thermopower. First-principle band structure shows that the chemical potential is close to the Dirac-cone-like points in linear bands. The magnetic field suppresses the apparent Hall carrier density of CaMnBi$_2$ below 50 K. The large magnetothermopower effect in (Sr/Ca)MnBi$_2$ is attributed to the magnetic field shift of chemical potential.

\end{abstract}


\maketitle 

The magnetic field influence on the thermal transport in ordinary metals is usually very small. Initially the large magnetothermopower effect was observed in doped InSb which was attributed to the effects of the sample geometry on the minority carriers.\cite{InSb} In a system with large magnetoresistant effect the magnetic field has significant influence on the properties of carriers and large magnetothermopower effect could be expected. The giant magnetothermopower effect was achieved in the giant magnetoresistant multilayer/granular systems and the colossal magnetoresistant manganites, which could be of interest for magnetic field sensors or magnetic controllable thermoelectric devices.\cite{gmr,cmr}

Ag$_{2-\delta}$Te and Ag$_{2-\delta}$Se are gapless semiconductors with linear energy dispersion. They were found to exhibit very large linear magnetoresistant effect.\cite{agte1,agte2,fangzhong} In materials with Dirac fermions the distance between the lowest and $1^{st}$ Landau level (LL) in magnetic field is very large and the quantum limit where all of the carriers occupy only the lowest LL is easily realized in relatively small fields. As a consequence, extraordinary quantum phenomena in magnetic field such as unsaturated magnetoresistance and Berry's phase were observed.\cite{graphene,ti} In Ag$_{2-\delta}$Te the magnetic field has significant influence on the thermopower as well as on the resistivity. The maximum value of magnetothermopower $S(9$ T$)-S(0) \sim 470$ $\mu$V/K at 110 K is observed. It was proposed that the Dirac carriers with linear energy dispersion play important role in the giant magnetothermopower effect.\cite{agtes} Recently, (Sr/Ca)MnBi$_2$ with two-dimensional Bi square nets are found to host Dirac fermions and the large linear magnetoresistance,\cite{srmnbi2,srmnbi22,kefeng1,kefeng2} but the magnetothermopower in these materials is still unknown.

Here we report the temperature and magnetic field dependence of the thermal transport properties in (Sr/Ca)MnBi$_2$ crystals with Dirac fermions. In SrMnBi$_2$, thermopower is positive indicating hole-type carriers. The magnetic field enhances the thermopower significantly. The maximum change of thermopower is about $1600\%$ in 9 T and at 10 K. In contrast, negative thermopower is observed in CaMnBi$_2$ with dominant electron-type carriers, whereas magnetic field suppresses the absolute value of thermopower. First-principle band structure shows that the chemical potential is close to the Dirac-cone-like points in linear bands. The giant magnetothermopower effects in bulk Dirac materials (Sr/Ca)MnBi$_2$ are attributed to the shift of chemical potential in the linear bands by the magnetic field.

Single crystals of SrMnBi$_2$ and CaMnBi$_2$ were grown using a self-flux method.\cite{kefeng1} The crystals are plate-like and the basal plane of a cleaved single crystal is the crystallographic $ab$-plane. The crystal was cleaved to a rectangular shape with dimension 5$\times$2 mm$^{2}$ in the \textit{ab}-plane and 0.2 mm thickness along the \textit{c}-axis for thermal transport measurement. Thermal transport properties were measured in Quantum Design PPMS-9 from 2 K to 350 K using one-heater-two-thermometer method. The direction of heat and electric current transport was along the $ab$-plane of single grain crystals with magnetic field along the \textit{c}-axis and perpendicular to the heat/electrical current. The relative error in our measurement was $\frac{\Delta \kappa}{\kappa}\sim$5$\%$ and $\frac{\Delta S}{S}\sim$5$\%$ based on Ni standard measured under identical conditions.  First principle electronic structure calculation were performed using experimental lattice parameters within the full-potential linearized augmented plane wave (LAPW) method~\cite{wien2k1} implemented in WIEN2k package.\cite{wien2k2}

\begin{figure}[tbp]
\includegraphics[scale=0.8]{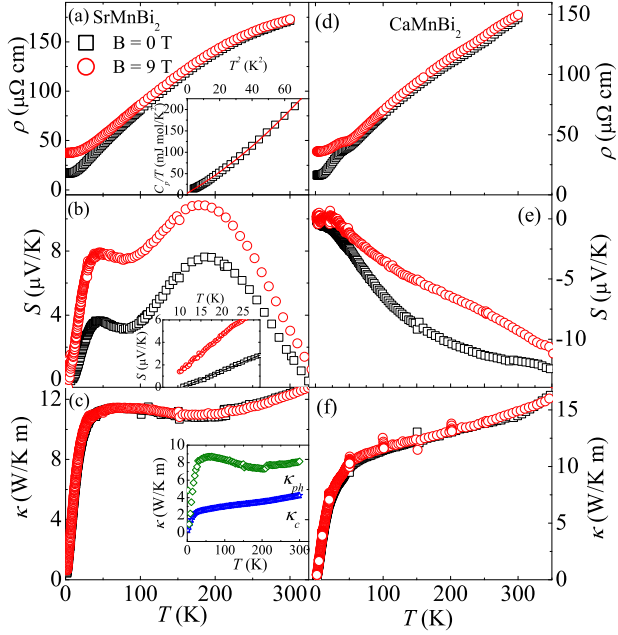}
\caption{(Color online) Resistivity $\rho$ (a,d) thermopower $S$ (b,e), and thermal conductivity $\kappa$ (c,f) for SrMnBi$_2$ (a-c) and CaMnBi$_2$ (d-f) in 0 T (squares) and 9 T (circles) magnetic field, respectively. The inset in (a) shows the specific heat data in the low temperature range as $C_p/T$ vs. $T^2$ and the line is the fitting results for SrMnBi$_2$. The inset in (b) shows the linear fitting results of $S$ in low temperature range. The inset in (c) shows the charge carrier ($\kappa_c$) and lattice thermal conductivity ($\kappa_{ph}$).}
\end{figure}

The in-plane resistivity $\rho(T)$ for SrMnBi$_2$ (Fig. 1(a)) exhibits a metallic behavior. An external magnetic field enhances the resistivity. As the temperature is increased, the magnetoresistance is gradually suppressed and becomes rather small above $\sim 60$ K. Thermopower $S$ for SrMnBi$_2$ is positive up to 340 K indicating hole-type carriers (Fig. 1(b)). Below 50 K, $S$ shows a nearly linear dependence with decreasing temperature in accordance with the diffusive thermopower response in metals.\cite{TEP1} In this temperature range, thermopower increases with the increase in temperature. There is a hump in thermopower curves at about 50 K, consistent with the thermal conductivity peak (Fig. 1(c)). The fitting of the specific heat data (inset of Fig. 1(a)) using $C_p(T)=\gamma T+\beta T^3+\eta T^5$, where $\gamma T$ is the residual Sommerfeld coefficient and $\beta T^3+\eta T^5$ is the phonon part of heat capacity, gives the Debye temperature $\Theta_D=310$ K. In the fitting it was necessary to include anharmonic phonon term $\sim T^5$. The hump in thermopower at $\sim 50$ K is approximately at $\Theta_D/5\sim 60$ K, implying that it originates from phonon drag effect.\cite{TEP1,TEP2}

At high temperature range, thermopower exhibits a peak at about 200 K and then decreases to negative values at about 350 K with further increase in temperature. External magnetic field has slight influence on the $S$ below $\sim 5$ K and above 250 K, but significantly enhances $S$ between 10 K and 200 K. The magnetic field has negligible influence on the thermal conductivity (Fig. 1(c)). The lattice thermal conductivity ($\kappa_{ph}$) is much larger than the carrier thermal conductivity ($\kappa_c$) which is derived from Wiedemann-Franz law $\kappa_c=L_0T/\rho$ with $L_0=2.44\times10^{-8}~W\Omega/K^2$ (inset of Fig. 1(c)). This implies that the phonon transport dominates the thermal conductivity but should not be dominant in the magnetothermopower. Results on CaMnBi$_2$ are similar (Fig. 1(d-f)).

\begin{figure}[tbp]
\includegraphics[scale=0.6]{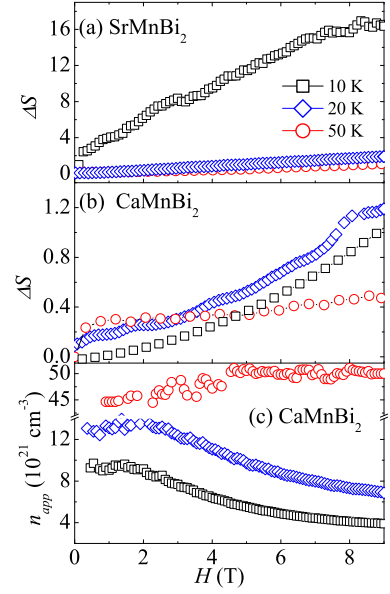}
\caption{(Color online) Magnetic field dependence of the magnetothermopower ratio $\Delta S=|(S(H)-S(0))/S(0)|$ for SrMnBi$_2$ (a) and CaMnBi$_2$ (b), as well as the apparent carrier density derived from Hall resistivity for CaMnBi$_2$ (c), at 10 K, 20 K, and 50 K, respectively. }
\end{figure}

The magnetic field dependence of the change in thermopower $S$ (magnetothermopower ratio $\Delta S=|(S(H)-S(0))/S(0)|$) for SrMnBi$_2$ at different temperatures is shown in Fig. 2(a). At 10 K, $\Delta S$ increases linearly with increase in magnetic field similar to the linear field dependent magnetoresistance ~\cite{kefeng1} and approaches $\sim 1600\%$ at 9 T. This value is comparable to values observed in managnites ($\sim 1400\%$) and Ag$_{2-\delta}$Te ($\sim 500\%$),\cite{cmr,agtes} although the value of $S$ is small (peaks at $\sim 10$ $\mu$V/K). With the increase in temperature, the magnetothermopower is suppressed but $\Delta S$ is still $\sim 200\%$ at 20 K and $\sim 150\%$ at 50 K.

\begin{figure}[tbp]
\includegraphics[scale=1]{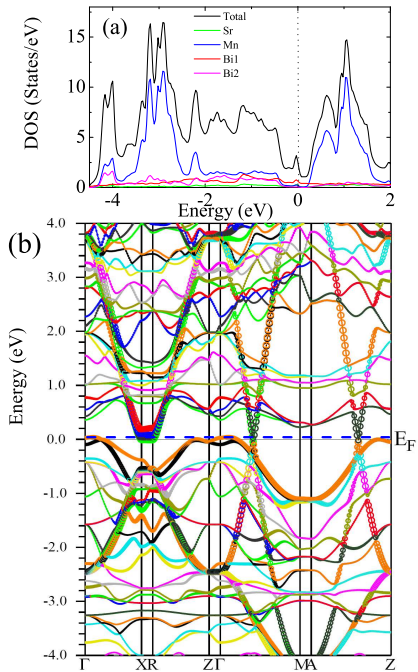}
\caption{(Color online) (a) The total density of states (DOS)(black line) and local DOS from Ca (blue line), Mn (green line), Bi square nets (Bi1, red line) and another type of Bi (Bi2, magenta line), for SrMnBi$_2$ respectively. (b) The band structure of for SrMnBi$_2$. The dashed line indicates the position of Fermi energy..}
\end{figure}

CaMnBi$_2$ has similar two dimensional Bi square nets but somewhat smaller unit cell since Ca ion has smaller radius than Sr ion. The magnetoresistant behavior of CaMnBi$_2$ is similar to SrMnBi$_2$ \cite{kefeng2} whereas the thermopower is different (Fig. 1 (d) and (e)). The thermopower $S$ is negative in the whole temperature range indicating electron-type carriers and the magnetic field suppresses the absolute value of $S$ (Fig. 1(e)). The magnetothermopower ratio $\Delta S$ is smaller than values in SrMnBi$_2$ and the maximum $\Delta S$ is about $120\%$ at 10 K and 9 T field (Fig. 2(b)). Fig. 2(c) shows the magnetic field dependence of the apparent Hall carrier density for CaMnBi$_2$, $n_{app}=H/(e\rho_H)$ at several different temperatures and in the simplest approximation where conductivity of a single carrier band is dominant. At low temperature $n_{app}$ is very sensitive to field. For example, the carrier density at 1 T is reduced up to half in magnitude in 9 T field. At $\sim$50 K, the magnetic field has no significant influence on the carrier density.

SrMnBi$_2$ and CaMnBi$_2$ host Dirac fermions with linear energy dispersion which dominate the electronic transport properties.\cite{srmnbi2,kefeng1,kefeng2} Both have small density of states at the Fermi level, as shown in Fig. 3(a). The first principle calculation indicates that the Fermi level in SrMnBi$_2$ is just below the Dirac-cone-like point in linear bands (Fig. 3(b)) and the dominant hole-type carriers.\cite{srmnbi22} This is consistent with the positive thermopower and nearly zero thermopower at low temperature due to near electron-hole symmetry in our measurement. CaMnBi$_2$ has similar structure and band structure, so the negative thermopower and nearly zero thermopower at 2 K implies that the Fermi level is just above the Dirac-cone-like point. A small change in the chemical potential at finite temperature will induce the significant change in the number and type of carriers, ultimately influencing the transport properties. The chemical potential at finite temperature depends on both the magnetic field and temperature.\cite{agtes} In CaMnBi$_2$, the decrease in temperature lowers the chemical potential which is just above the Dirac-cone-like point toward this point, and then induces the decrease in number of carriers (electrons). The magnetic field suppression of the apparent Hall carrier density in Fig. 2(c) demonstrates this point. Consequently the absolute value of thermopower is suppressed with decrease in temperature. The magnetic field also lowers the chemical potential and suppresses the thermopower.

In SrMnBi$_2$, the magnetic field will lower the chemical potential away from the Dirac-cone-like point and will induce the increase in the number of carriers (holes). This will ultimately enhance the thermopower. The thermopower of SrMnBi$_2$ below 30 K in zero and 9 T field shows a linear temperature dependence (inset in Fig. 1(b)), indicating that the diffusion mechanism dominates. Diffusive Seebeck response of a Fermi liquid is expected to be $S=\pm\frac{\pi^{2}}{2}\frac{k_B^2}{e}\frac{T}{\varepsilon_F}$ where $\varepsilon_F$ is the Fermi energy.\cite{TEP1,TEP2} Both the thermopower below 30 K in 0 T and 9 T field can be described by this formula very well (lines in the inset of Fig. 1(b)), but in 9 T field the slope of the low-temperature thermopower is much higher than the one in 0 T field. The higher slope implies smaller Fermi energy, so the change of the slope of the linear thermpower in Field indicates the lowering of the chemical potential. Based on above discussion, the doping with holes in SrMnBi$_2$ can further lower the chemical potential and significantly enhanced thermopower could be expected.

In summary, we report large magnetothermopower effect in the recently reported bulk Dirac materials (Sr/Ca)MnBi$_2$. In SrMnBi$_2$, thermopower is positive indicating hole-type carriers. The maximum change of thermopower is about $1600\%$ in 9 T and at 10 K. A negative thermopower is observed in CaMnBi$_2$ with electron-type carriers whereas magnetic field suppresses the magnitude of thermopower. In both cases, the magnetic field has no influence on the thermal conductivity. The large magnetothermopower effects in Dirac materials (Sr/Ca)MnBi$_2$ are attributed to the shift of chemical potential in the linear bands by magnetic field.

~\\
We thank John Warren for help with SEM measurements. Work at Brookhaven is supported by the U.S. DOE under contract No. DE-AC02-98CH10886.


\end{document}